%% file: 3-c-springer.tex
\documentclass[runningheads]{llncs}
\usepackage{graphicx}
\usepackage{booktabs}
\usepackage{amsmath,amssymb,amsfonts}
\usepackage{algorithmic}
\usepackage{textcomp}
\usepackage{xcolor}
\usepackage{booktabs}
\usepackage{lipsum,lineno}

\input{0-imports}

\input{0-definitions}

\begin{document}

\title{\mytitle\thanks{We thank our colleagues (N.A., J.W., A.B., R.B., C.C., C.L., P.M., S.A.) for their support, and the students who generously volunteered to participate in the study.}
}
\titlerunning{\myshorttitle}
%

\ifdefined\Anonymous 

    \author{Anonymous 1\inst{1,2} \and
    Anonymous 2\inst{2} \and
    Anonymous 3\inst{1} \and
    Anonymous 4\inst{1} \and
    Anonymous 5\inst{2}
    }
    \authorrunning{Anonymous 1 et al.}
    %
    \institute{Affiliation 1
    Affiliation 2
    }

\else

    \author{\AuthorJB\inst{1,2}\orcidID{\ORCIDJB} \and
    \AuthorLH\inst{2}\orcidID{\ORCIDLH} \and
    \AuthorKU\inst{1}\orcidID{\ORCIDKU} \and
    \AuthorFM\inst{1}\orcidID{\ORCIDFM} \and
    \AuthorEB\inst{2}\orcidID{\ORCIDEB} 
    }
    \authorrunning{Brender et al.}
    %
    \institute{\MOBOTS \\ \email{\EmailJB, \EmailKU,  \EmailFM} \and
    \HEPVaudEnglish \\
    \email{\EmailEB,\EmailLH}
    }
    
\fi

\maketitle

\begin{abstract}
    \input{1-abstract}

\end{abstract}

\input{2-article}

\vspace{-10pt}
\bibliographystyle{splncs04}
\bibliography{0-bib}

\end{document}

%% file: 0-imports.tex
\usepackage{todonotes}
\usepackage{svg}
\usepackage{array}\newcolumntype{P}[1]{>{\centering\arraybackslash}p{#1}}
\usepackage{ifthen}
\usepackage{xcolor}
\usepackage{cite}
\usepackage{verbatim}
\usepackage{placeins}
\usepackage{lscape}
\usepackage{multirow}
\usepackage{graphicx}
\usepackage{float}
\usepackage{lipsum}  
\usepackage{todonotes}
\usepackage{nicematrix}
\usepackage{hyperref}
\usepackage{enumitem}
\usepackage{siunitx}
\usepackage{graphicx}
\usepackage{subcaption}  
\usepackage{adjustbox}

\definecolor{lavenderblush}{rgb}{0.93, 0.88, 0.90}

\colorlet{shadecolor}{lavenderblush}

%% file: 0-definitions.tex
\newcommand{\mytitle}{Structured Prompts, Better Outcomes? Exploring the Effects of a Structured Interface with ChatGPT in a Graduate Robotics Course}
\newcommand{\myshorttitle}{Structured Prompts, Better Outcomes?}

\newboolean{ISanonymous}
\setboolean{ISanonymous}{False}
\ifthenelse{\boolean{ISanonymous}}{\newcommand*{\Anonymous}{}}

\newcommand{\KWA}{ChatGPT}
\newcommand{\KWB}{Computing Education}
\newcommand{\KWC}{Behaviors}
\newcommand{\KWD}{Performance}
\newcommand{\KWE}{Learning}
\newcommand{\KWF}{Effectiveness}

\newcommand{\MOBOTS}{MOBOTS \& LEARN, EPFL, Switzerland}

\newcommand{\HEPVaudEnglish}{University of Teacher Education (Haute École Pédagogique) Vaud, Switzerland}

\newcommand{\AuthorLH}{Laila El-Hamamsy}
\newcommand{\AuthorKU}{Kim Uittenhove}

\newcommand{\AuthorFM}{Francesco Mondada}

\newcommand{\AuthorEB}{Engin Bumbacher}

\newcommand{\AuthorJB}{Jérôme Brender}

\newcommand{\EmailLH}{laila.elhamamsy@hepl.ch}
\newcommand{\EmailKU}{kim.uittenhove@epfl.ch}
\newcommand{\EmailEB}{engin.bumbacher@hepl.ch}

\newcommand{\EmailJB}{jerome.brender@epfl.ch}

\newcommand{\EmailFM}{francesco.mondada@epfl.ch}

\newcommand{\ORCIDLH}{0000-0002-6046-4822}
\newcommand{\ORCIDKU}{0000-0001-5450-3875}

\newcommand{\ORCIDEB}{0000-0002-4322-7059}

\newcommand{\ORCIDFM}{0000-0001-8641-8704}

\newcommand{\ORCIDJB}{0009-0008-3279-5641}

%% file: 1-abstract.tex
Prior research shows that how students engage with Large Language Models (LLMs) influences their problem-solving and understanding, reinforcing the need to support productive LLM-uses that promote learning. This study evaluates the impact of a structured GPT platform designed to promote ``good'' prompting behavior with data from 58 students in a graduate-level robotics course. The students were assigned to either an intervention group using the structured platform or a control group using ChatGPT freely for two practice lab sessions, before a third session where all students could freely use ChatGPT.
We analyzed student perception (pre-post surveys), prompting behavior (logs), performance (task scores), and learning (pre-post tests). Although we found no differences in performance or learning between groups, we identified prompting behaviors - such as having clear prompts focused on understanding code - that were linked with higher learning gains and were more prominent when students used the structured platform. However, such behaviors did not transfer once students were no longer constrained to use the structured platform. Qualitative survey data showed mixed perceptions: some students perceived the value of the structured platform, but most did not perceive its relevance and resisted changing their habits. 
These findings contribute to ongoing efforts to identify effective strategies for integrating LLMs into learning and question the effectiveness of bottom-up approaches that temporarily alter user interfaces to influence students' interaction. Future research could instead explore top-down strategies that address students’ motivations and explicitly demonstrate how certain interaction patterns support learning.

%% file: 2-article.tex

\section{Introduction and Related Work}
Since the public release of ChatGPT in late 2022, large language models (LLMs) have gained rapid adoption for a wide range of tasks and fields \cite{singhal2023large,li2022competition, noy2023experimental}. Early studies show that access to such tools can significantly improve productivity in knowledge-intensive work \cite{noy2023experimental, singhal2023large}. 
Although a recent meta-analysis of 51 experimental studies across multiple disciplines found ChatGPT use to generally enhances academic performance \cite{deng2024does}, other work has found that students can use ChatGPT in ways detrimental for learning \cite{brender2024s}, that use of ChatGPT can lead to procrastination \cite{swargiary_impact_2024} or laziness \cite{yilmaz2023augmented}, to increased memory retention issues \cite{abbas2024harmful}, and to reduced academic performance \cite{ abbas2024harmful}. Such findings demonstrate that many questions remain about the impact of LLMs on learning.


\vspace{-5pt}
\subsection{Over-reliance on ChatGPT in computing education contexts}
\vspace{-5pt}

In computing education, understanding the impact of LLMs, such as ChatGPT, is critical as AI-based code generation tools are increasingly adopted by students and teachers \cite{bahroun2023transforming}. 
On the one hand, LLMs can scale the benefits of one-on-one tutoring \cite{bowman2013academic}, aid conceptual understanding, and expand topic coverage in programming contexts \cite{liu2024teaching}. 
On the other hand, concerns remain that LLMs may inhibit learning by reducing cognitive effort, especially in domains requiring active problem-solving, such as programming. For example, in mathematics, Bastani et al. \cite{bastani2024generative} show that overreliance on LLMs leads to a decline in performance when LLM support is removed, highlighting the need for LLM-interfaces that support learning and ``ensure that humans continue to learn critical skills'' \cite{bastani2024generative} and prevent ``metacognitive laziness'' \cite{fan2025beware}. 

More recent studies suggest a nuanced picture: when students substitute their learning activities with LLMs (e.g., by generating complete solutions to exercises), they can cover a larger breadth of topics but understand them less deeply \cite{lehmann2025aimeetsclassroomlarge}. In contrast, students who complement their learning with LLMs—using them to reflect or clarify rather than replace their work—show improved conceptual understanding without necessarily expanding topic coverage \cite{lehmann2025aimeetsclassroomlarge}. Finally, another study demonstrated that while students may improve their performance while using ChatGPT, this does not necessarily translate into improved conceptual understanding \cite{brender2024s}, with the question of how ChatGPT is used being of the utmost importance. 
These findings thus highlight the importance of understanding the impact of how students use LLMs and how specific types of LLM use impact learning \cite{brender2024s}. Indeed, while thoughtful engagement with LLMs can foster learning gains, excessive reliance on these tools, especially for solving practice exercises, may impair learning outcomes \cite{lehmann2025aimeetsclassroomlarge}.

These findings underline the importance of both examining how students use LLMs and teaching them how to use LLMs effectively. 

\vspace{-5pt}
\subsection{The importance of establishing guidelines for effective LLM use}
\vspace{-5pt}

Crafting effective prompts remains challenging, particularly for users without LLM expertise \cite{abdellatif2020challenges}. Prompt engineering has therefore emerged as a strategy to improve interactions with LLMs, alongside efforts to formalize prompting with guidelines \cite{denny2023promptly, mollick2022new, stokel2023chatgpt}. Yet these guidelines often lack empirical grounding and rarely address how prompting supports understanding and transfer to other environments with LLMs. To foster LLM literacy, especially among students, there is a need to translate such guidelines into accessible, evidence-informed formats \cite{yang2023use} and ensure that changes in practices persist after guidance is removed (i.e., transfer). 
Some interventions have introduced AI tutors \cite{favero2024enhancing, bastani2024generative} that personalize student interactions with LLMs, but these approaches raise even more concerns about dependency on LLMs. One study found that once scaffolded support provided by an AI tutor is removed, students struggle to engage meaningfully with the AI on their own, potentially impairing long-term skill development \cite{bastani2024generative}.  
It therefore appears that the transfer of productive prompting strategies from guided to unguided AI environments remains poorly understood. 

\vspace{-5pt}
\subsection{Motivation}
\vspace{-5pt}


To address this gap, we designed a structured, form-based interface to structure students' prompting of ChatGPT during programming labs at the university level (see section \ref{sec:BOMR}). The interface decomposed prompt construction into blank-based fields, encouraging more reflective and deliberate AI use. Unlike Socratic bots \cite{favero2024enhancing} that aim to simulate pedagogical dialogue, our goal was to empower students to develop their own prompting strategies.
With this structured interface, we sought to answer the following research question: How does exposure to a structured prompting interface impact good prompting practices, student learning and performance during practice labs?
To answer this question and investigate (i) whether guided prompting habits persist once support is withdrawn (i.e., transfer), and (ii) whether such structuring contributes to improved learning and performance, we conducted an intervention over three lab sessions. During the first two sessions the students were divided into two groups: those with access to ChatGPT, and those with access to the structured interface. In the third session, the structured interface was removed, and students were allowed to interact freely (i.e., without constraint) with ChatGPT. 
For the purpose of the present article, we focused our analysis on the data from the second practice lab session, where half the students were required to use the structured interface, and the third practice lab session, where both groups could freely use ChatGPT.

\vspace{-5pt}
\section{The structured interface with ChatGPT API}
\label{sec:BOMR}
\vspace{-5pt}

\begin{figure}[!htpb]
    \centering
    \includegraphics[width=0.8\linewidth]{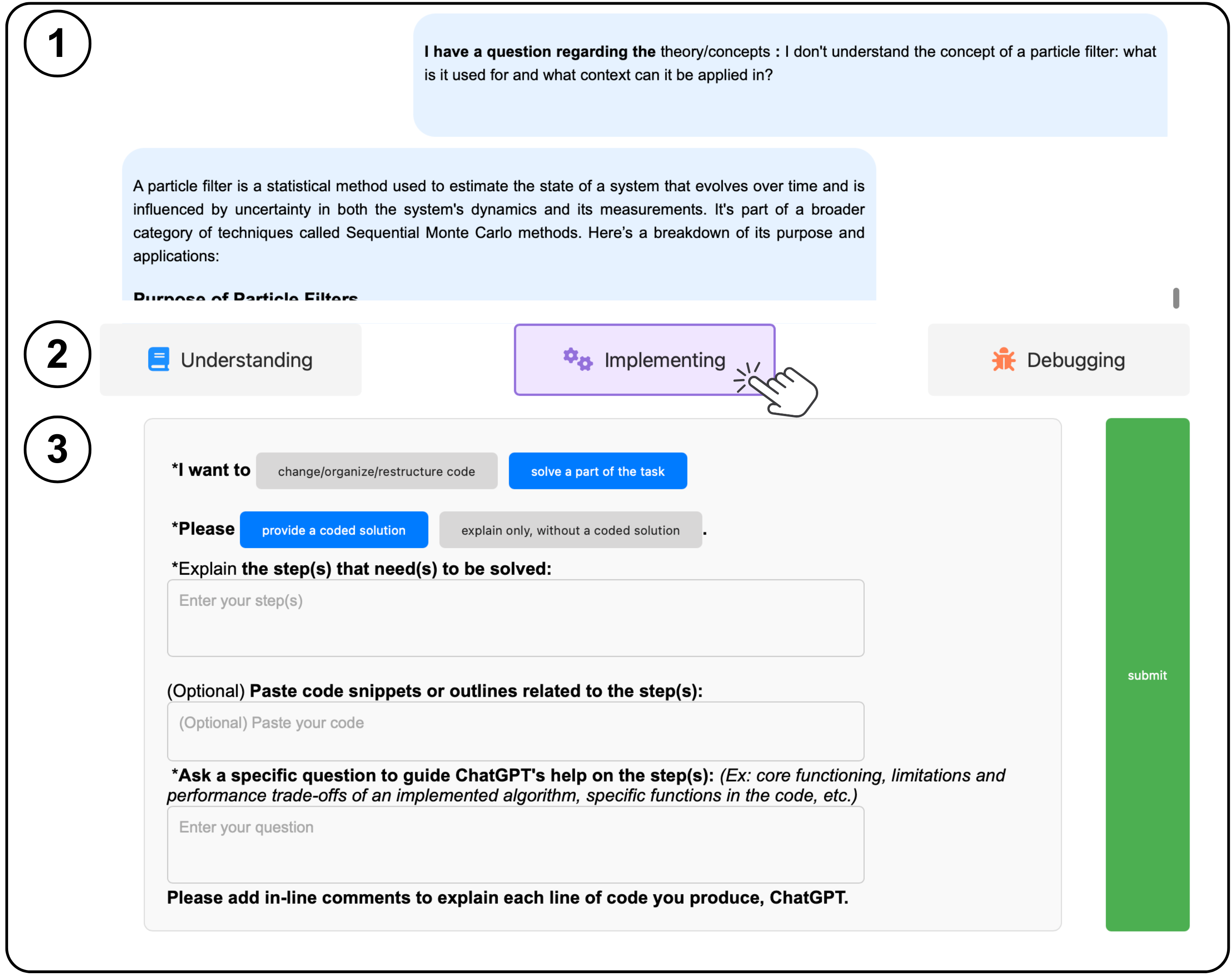}
    \caption{Structured interface with integrated GPT API comprising of (1) an interactive interface, (2) prompt type category selection (understanding, implementing, or debugging), and (3) a structured format for prompt composition.}
    \vspace{-15pt}
    \label{fig:platform}
\end{figure}

Drawing from research on metacognitive scaffolding and reflective learning \cite{byun2014relative,prather2019first}, we developed a structured interface (see Fig. \ref{fig:platform}) that acts as a reflective layer, encouraging students to clarify their intent before submitting a prompt. A similar approach has previously demonstrated effectiveness in promoting metacognitive awareness among novice programmers \cite{loksa2016programming}. Concretely, the structured interface was designed to help students decompose their prompts on GPT-based tools. Students must first select a prompt category drawing from prior research on prompt classification in similar educational contexts \ifdefined\Anonymous \texttt{[Anon.]}\else \cite{brender2024s}\fi (\textit{Understanding}, \textit{Implementing}, or \textit{Debugging}, see item 2 in Fig.~\ref{fig:platform}). This categorization aimed to raise students’ awareness of their goals and reasoning when engaging with the system.

Once students select the prompt category, a corresponding structured form appears (see Fig.~\ref{fig:platform} item 3) to have students elaborate on their query
. These forms were informed by research on effective questioning in ill-structured problem-solving \cite{byun2014relative} and 
programming \cite{loksa2016programming,prather2019first} to encourage metacognitive engagement, prompting learners to reflect on and refine their own prompts. Once completed, the structured fields are compiled into a single prompt and sent to the ChatGPT API (GPT-o-mini, free access at the time of the study). The interface did not alter or constrain the underlying knowledge base of the GPT model.

\vspace{-5pt}
\section{Methodology}

\subsection{Study Context and Design}
\vspace{-5pt}

A mixed-methods classroom intervention was conducted at the university level\footnote{The study was approved by EPFL's Ethics Committee \ifdefined\Anonymous\else (HREC000560/12.08.2024) \fi and students' informed consent was obtained prior to participation.} 
to address our RQs and understand how engaging first with a structured GPT platform, and subsequently with GPT alone, affects students' performance, learning outcomes, and perceptions during a practical lab session.

\vspace{-5pt}
\subsubsection{Participants}

The study took place during the practice lab sessions of a graduate-level mobile robotics course at \ifdefined\Anonymous \texttt{Anon. university}\else EPFL\fi and involved 58 (39 male, 18 female, 1 undisclosed) of 143 students. These students volunteered and consented to participate in the study, in exchange for monetary compensation.

\begin{figure}
    \centering
    \includegraphics[width=1\linewidth]{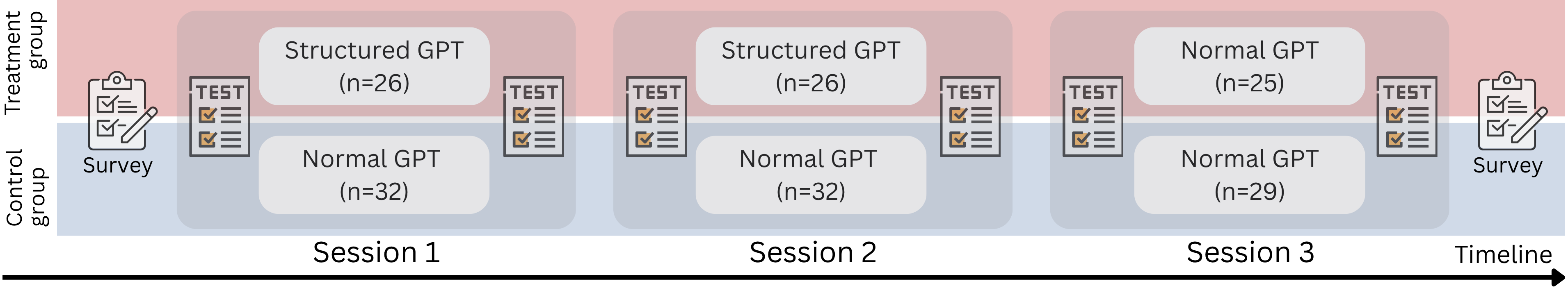}
    \caption{Overview of study design with participants allocated to structured GPT and normal GPT interfaces. A pre-survey was conducted before practice lab session 1 and a post-survey after session 3. Learning assessments (pre/post-tests), prompt logs, and task performance logs were collected each session.}
    \label{fig_study_design}
\end{figure}

\vspace{-7pt}
\subsubsection{Study Design}

The study outline is reported in Fig. \ref{fig_study_design}.
Before the first session, a preliminary survey on students' prior LLM perception and use was administered. 

The 58 students were then randomly assigned into two groups: an intervention group using the structured interface (n=26) and a control group using ChatGPT freely (n=32). Please note that we ensured that both groups were equivalent in terms of 
(i) prior usage and perception based on the first perception survey (Mann Whitney U test, $p>0.05$ on all questions) 
(ii) prior knowledge based on each session's pre-test (Mann-Whitney, session 2: $U=454.0, p=.55, M=60\pm26\%$, session 3: $U=387.0, p=.87, M=44.1\pm27.7\%$)

During the first two sessions the students
were divided into two groups: those with access to ChatGPT and those with access to the structured interface. In the third session, all the students interacted freely with ChatGPT.

Finally, the students responded to a post-survey to establish whether the intervention led to changes in LLM perception and use. 

\vspace{-7pt}
\subsubsection{Practice Lab Session}

Each practice lab session was structured as follows: 

(i) A 15 minute briefing on the study objectives, including a 10 minutes pre-test assessing prior knowledge of the course material. 

(ii) 75 minutes of hands-on tasks\footnote{The hands-on tasks are part of a robotics course and focused on the Dijkstra and A-star algorithms in session 2, and particle filters in session 3. All tasks were computer-based: students coded robotics algorithms without using physical robots.} that are decomposed into subtasks (5 in Session 2 and 8 in Session 3) graded on a 0, 0.5, or 1 scale. The complexity of the tasks made the use of ChatGPT pedagogically meaningful.

(iii) A 10 minute post-test with in-depth questions on the same content. 

\vspace{-7pt}
\subsection{Data Collection and Reliability Measures}

\subsubsection{Perception Surveys} 
The pre and post intervention surveys were adapted from the Technology Acceptance Model (TAM) developed for educational contexts \cite{teo2009modelling}, and encompasses five dimensions: utility, ease of use, attitude/interest, self-efficacy, and intent to employ in future teaching activities\footnote{The complete survey is accessible \href{https://drive.switch.ch/index.php/s/WEWpXIgP4E5jjcO}{here}.}. 

\vspace{-5pt}
\subsubsection{Pre-Test}
Each session included a pre-test with two questions: (1) explain the algorithm (3 points); (2) sequence the steps of the algorithm (3 points). Open-ended responses were anonymously graded by two Teaching Assistants (TAs) who reached substantial inter-rater reliability on the 20\% of the data (IRR, $\kappa_{S2}=0.71$, $\kappa_{S3}=0.75$) before coding the full dataset  ($\kappa_{S2}=0.69$, $\kappa_{S3}=0.71$). 

\vspace{-5pt}
\subsubsection{Practice Lab Tasks (Performance)}
The codes produced for each task (5 in session 2, 8 in session 3) were graded 
by three TAs using a 3-point scale: $0$ for incomplete, $0.5$ for over 70\% correct, and $1$ for fully correct solutions. We randomly selected $20\%$ of the codes of each task and had the three TAs grade them and achieved substantial agreement on each task (Fleiss’ $\kappa_{S2}$ in $[0.71; 0.77]$; and $\kappa_{S3}$ in  $[0.7,0.78]$). The remaining data was distributed between the three TAs and coded individually. The total practice lab score is computed as the standardized sum of the individual tasks' scores.

\vspace{-10pt}
\subsubsection{ChatGPT Prompts (Usage)}
We analyzed 291 prompts from session 2, and 411 from session 3\footnote{Excluding 16 and 18 off-task prompts (e.g., ``hello'', jokes) from sessions 2 and 3.} from 58 students (32 control, 26 intervention) who consented to share them ($M_{S2} =4.84\pm3.97$ $M_{S3}=7.61\pm4.23$ per student). Two researchers first analyzed a subset to define a prompt categorization based on observed uses of ChatGPT in computing education \ifdefined\Anonymous \texttt{\cite{lau_ban_2023} and \texttt{[Anon.]}} \else \cite{lau_ban_2023, brender2024s}\fi, resulting in three main prompt types (as in \cite{brender2024s}): (i) \emph{Development}, i.e. generating code or solutions; (ii) \emph{Conceptual}, i.e. explaining task elements or computing concepts; (iii) \emph{Debugging}, i.e. identifying or resolving coding errors. Three TAs obtained substantial agreement ($\kappa=.71$) on 20\% of the dataset, before coding the full dataset.

To provide more insight into students' prompts, three additional binary attributes were annotated per prompt 
whether the student was seeking to increase their understanding (\emph{Understanding}, e.g., ``explain this code [...]''), whether student was specific and provided details in their prompt (\emph{Granularity}, e.g., referring to variables or lines of code rather than general tasks), whether the student had clear, well-structured, unambiguous phrasing (\emph{Clarity}). Three TAs obtained substantial agreement on 20\% of the data ($\kappa_{understanding}=.76$, $\kappa_{granularity}=.74$, $\kappa_{clarity}=.68$), before coding the remaining prompts individually. 

\begin{table}[ht]
\centering
\caption{Prompt coding explanations, examples and counter-examples}
\label{tab:Prompt_Quality_Dimensions}
\begin{adjustbox}{center}
\begin{tabular}{l p{4cm} p{4.5cm} p{4cm}}
\toprule
Dimension & Explanation & Example included& Counter example not included\\
\midrule
Understanding & Clarifies or seeks to understand concepts explicitly. & ```explain me this code [...])'' & ``Here's my code.'' \\
Granularity  & Specificity of the prompt regarding a part of the task or code (fine-grained). & ``Could you explain line 4, specifically why the index is set to zero initially?'' & ``What does this function do?'' \\
Clarity& Clear and explicit request structure. & ``I started coding the algo and need help debug this error:[...], explain me the bug.'' & ``Here my bug:'' \\
\bottomrule
\end{tabular}
\end{adjustbox}
\end{table}

Finally, to quantify the proportion of personal input within each prompt, two teaching assistants (TAs) collaboratively worked on identifying where the student's original contribution began and ended. 
Based on these annotations, we computed the percentage of each prompt that consisted of personally written input relative to the total prompt length.

\vspace{-10pt}
\subsubsection{Post-Test (Learning)}
The post-tests were in MCQ format and consisted of 8 advanced questions in session 2 and 5 in session 3, targeting conceptual understanding of the practice lab session's subject. 
One point was attributed per question, and then summed and standardized to compute the overall post-test score per session ($M_{S2}=66\pm18\%$; $M_{S3}=55\pm20\%$).

\vspace{-5pt}
\subsection{Data Analysis}

\subsubsection{Effect of the structured intervention on learning and performance}
For sessions 2 and 3, we used multiple linear regressions and continuous variables so that the resulting standardized regression coefficients $\beta$ provide effect size estimates. Each model includes (1) the group (control, treatment) and (2) the interaction between the group and pre-test score to see how prior knowledge might modify the intervention's effect. In addition, we computed a normalized learning gain $ \frac{\text{post} - \text{pre}}{100\% -\text{pre}}$ to capture individual learning improvement \cite{nissen2018comparison}, with the exclusion of two students with full scores in the pre-test.
To evaluate whether certain prompting behaviors were associated with higher performance, we conducted a forward stepwise regression and then a multiple linear regression on performance (post-test and practice lab), with prompt types as predictors.
\vspace{-10pt}
\subsubsection{Effect of the structured intervention on ChatGPT usage} 

For sessions 2 and 3, we compared the prompt types, attributes and amount generated by the students in both conditions using t-tests, $\chi^2$ tests, and Mann-Whitney U tests. 

\vspace{-10pt}
\subsubsection{Students' Perception of ChatGPT usage} Having established that there were no pre-existing differences between students' perceptions of ChatGPT useage prior to the intervention, we used non-parametric Mann-Whitney U tests to determine whether there were differences between conditions after the intervention, and provide Cohen's d effect size when the difference is significant.

\vspace{-5pt}
\section{Results}

\subsection{Effect of the structured intervention on learning \& performance}
\label{Effect_of_the_structured}
\vspace{-5pt}

For sessions 2 and 3, we implemented multiple linear regression models with (i) post-test scores (learning), and (ii) practice lab scores (performance) as dependent variables and (a) pre-test scores, (b) group (control vs. intervention) and (c) their interaction as independent variables. None of the models revealed a significant main effect of the group, nor a significant interaction effect between group and pre-test scores on either performance or learning. The only significant predictor of learning was pre-test scores in session 3 ($\beta=0.521$; p=$.007$)
. 

An additional analysis of task completion rates during the practice labs indicates that there is no difference between groups in terms of (i) the number of completed tasks (t-test, session 2: $t(57)=0.23$, $p=.81$, session 3: $t(53)=0.74$, $p=.46$), and (ii) the number of attempted tasks (t-test, session 2: $t(57)=0.02$, $p=.97$, session 3: $t(53)=0.68$, $p=.50$).

Therefore, it appears that there is no direct impact of the intervention on learning or performance. To understand this finding, we investigated whether certain prompting behaviors were more conducive to learning, and whether using the structured interface could have promoted such behaviors, but maybe not enough as to be reflected in learning or performance differences.

\vspace{-5pt}
\subsubsection{The Influence of Specific Prompting Patterns on Learning:}


As will be explained in the next sub-section, we did not find significant differences between conditions in the use of prompt types (``Conceptual'', ``Debugging'', ``Development''), nor in the characteristics of prompts when examined separately (``Understanding'', ``Clarity'', ``Granularity''). Thus, we decided that for this analysis we could aggregate the two conditions into one dataset and examine whether overall behaviors emerged that seemed productive for learning.

As \textit{Conceptual} prompts and \textit{Debugging} prompts account for less than 15\% of all prompts, we focused on \textit{Development} prompts ($>$70\% of prompts) in order to examine the relationship between prompting behaviors and learning. 

A stepwise forward regression was conducted using \textit{Development}, \textit{Clarity}, \textit{Understanding}, and \textit{Granularity}, and their interaction as predictors of learning gain. For the session 2, the final model retained only \textit{Clarity}; however, the overall model was not statistically significant, $F(1, 56)=2.06$, $p=.15$, explaining only 3.6\% of the variance in \textit{learning\_gain} ($R^2=.036$). However, for the session 3, the final model retained all the variables.
The model explains approximately 45\% of the variance in learning gains ($R^2=.45$; adjusted $R^2=.35$) and the overall model was statistically significant ($F(8, 45)=2.06$, $p<0.01$).

\begin{table}[!ht]
\centering
\caption{Regression model for the impact of specific development-related prompting patterns on learning}
\label{tab:regression_table}
\begin{tabular}{lccc}
\toprule
& \multicolumn{3}{c}{Learning Gain} \\
\cmidrule{2-4}
Predictors & Estimate & CI & \textit{p}\\
\midrule
(Intercept)                & 0.19**   & 0.06 -- 0.31   & 0.005    \\
development                & 0.13     & -0.06 -- 0.32  & 0.189    \\
clarity                    & 0.05     & -0.13 -- 0.23  & 0.590    \\
understanding              & 0.23*    &  0.05 -- 0.43  & 0.017    \\
granularity                & -0.04    & -0.22 -- 0.14  & 0.685    \\
development:clarity        & 0.77***  &  0.48 -- 1.06  & 5e-06    \\
clarity:understanding      & 0.75***  &  0.45 -- 1.05  & 1e-05    \\
development:granularity    & -0.29$^\dagger$ & -0.58 -- 0.01 & 0.065    \\
understanding:granularity  & -0.36*   & -0.66 -- -0.06 & 0.022    \\
\midrule
Observations \ \ \ \ \ \ \  \  \ \ 54 & \multicolumn{3}{c}{}
$R^2$ / $R^2$ adjusted \ \ \ \ {0.45 / 0.35} \\
\bottomrule
\multicolumn{4}{c}{$^\dagger$\textit{p}$\le0.1$ \quad *$\textit{p}\le\textit{0.05}$ \quad **$p\le\textit{0.01}$ \quad ***$p\le\textit{0.001}$}  \\
\end{tabular}
\end{table}

The results show that a larger proportion of \emph{Understanding} prompts is a significant positive predictor ($\beta=0.23$, $p=0.017$, $p<0.05$) of learning gain. 
Although the main effect of having clear prompts is not significant (\emph{Clarity} $\beta=0.05$, $p=.58$), having development prompts that are clear (interaction between \emph{Clarity} and \emph{Development}, $\beta=0.77$, $p<.001$) and clear prompts that seek to improve students' understanding (interaction effect between \emph{Clarity} and \emph{Understanding}, $\beta=0.75$, $p<.001$) contribute to learning. These findings suggest that higher \emph{Clarity} amplifies the positive relationships of \emph{Development} and \emph{Understanding} with learning gains.
A significant negative interaction between \textit{Understanding} and \textit{Granularity} ($\beta=-0.36$, $p=.022$) suggests that overly detailed prompts may dampen the benefits of learning. However, no significant interaction was found between \textit{Development} and \textit{Granularity} ($\beta=-0.29$, $p=.29$).

In summary, \emph{Understanding} enhances learning, and its effects are substantially strengthened when \emph{Clarity} is high, and when Clarity is associated to \textit{Development} prompts. When \emph{Clarity} is low, their positive influences are comparatively weaker.
To understand why there were no significant learning or performance differences between conditions, we looked at both groups' usage patterns.

\vspace{-5pt}
\subsection{Effect of the structured intervention on ChatGPT usage:}

\vspace{-5pt}
\subsubsection{Number of prompts}
On average, students submitted $4.8\pm4.0$ prompts in session 2 and $7.6\pm4.2$ prompts in session 3, with no significant differences between conditions (t-test  $t(57)_{S2}=1.27$, $p_{S2}=0.2$ , and $t(53)_{S3}=0.17$, $p_{S3}=0.86$). However, the group using the structured interface tended to disengage more (46\% did $\leq 3$ prompts, compared to 25\% in the control group).

\vspace{-10pt}
\subsubsection{Types of prompts:}
There are no significant differences between conditions for the type of prompts (Conceptual: $U=470$, $p=.14$; Development: $U=386$, $p=.95$; Debugging: $U=321$, $p=.19$). 
In both sessions, prompts categorized as \textit{Development} prompts were the most frequent, accounting for over 70\% of students' prompts in both sessions\footnote{Average proportion of prompts: Development session 2: $70.0\pm31.5 \%$, session 3: $72.7\pm26.2 \%$; Conceptual session 2: $11.1\pm24.6 \%$, session 3: $14.2\pm23.5 \%$; Debugging session2: $12.5\pm22.8 \%$, session 3: $9.4\pm15.2 \%$.}. 
In contrast, \textit{Conceptual} and \textit{Debugging} prompts were less common, accounting for approximately 10\% of students' prompts, with many students submitting no \textit{Conceptual} or \textit{Debugging} prompts\footnote{44 students in session 2 and 31 in session 3 submitted no conceptual prompts. 38 students in session 2 and 34 in session 3 submitted no debugging prompts.
}

\vspace{-10pt}
\subsubsection{Prompt attributes:}
On average, students in both conditions had prompts with similar levels of \textit{Understanding} (i.e., prompts with questions to improve understanding) (37.1\% in session 2 and 23.3\% in session 3), \textit{Granularity} (i.e., prompts with sufficient detail provided) (21.2\% in session 2 and 20.4\% in session 3), and \textit{Clarity} (i.e., prompts with a clear and explicit request) (37.5\% in session 2 and 16.5\% in session 3).
There were no significant differences between conditions.


We extend our analysis by focusing on \textit{Development} prompts and their interaction with attributes, in light of the learning effects identified in Section~\ref{Effect_of_the_structured} for both session 2 and 3.
It appears that students in the intervention group had a higher proportion of clear development prompts (prompts coded as \textit{Development} and \textit{Clarity}), $\chi^2$(3)=8.57, $p=.036$, in session 2, when using the structured interface, but there was no significant difference in session 3 when all students could use the normal ChatGPT interface. As a reminder (see Table \ref{tab:regression_table}), a higher proportion of such types of prompts are significant positive predictors of learning. No other associations between \textit{Development} prompts and quality dimensions were statistically significant between groups (see Table~\ref{chi_square}). We also examined condition differences for both sessions for the other combinations of other prompt categories and attributes, but found no significant differences.

This finding indicates that the structured interface seems to have been conducive to the increased use of Clear \textit{Development} prompts, which are associated with higher learning gains, but that this effect was short-lived: When the structured interface was removed (session 3), students' use of Clear Development prompts was not different from the comparison group.

\vspace{-15pt}
\begin{table}[ht]
\centering
\caption{Chi-square test comparing the joint distribution of prompt types and metrics between conditions for sessions 2 and 3.}
\begin{tabular}{llcccccc}
\hline
\textbf{Category} & \textbf{Attributes} & \multicolumn{3}{c}{\textbf{Session 2}} & \multicolumn{3}{c}{\textbf{Session 3}} \\ \cline{3-8} 
                  &                  & \(\chi^2\)     & p-value     & dof     & \(\chi^2\)     & p-value     & dof     \\ \hline

Development       & Understanding    & 0.92           & 0.820       & 3       & 0.96           & 0.812       & 3       \\
Development       & Granularity      & 2.26           & 0.521       & 3       & 3.21           & 0.360       & 3       \\
Development       & Clarity          & 8.57           & 0.036       & 3       & 4.69           & 0.196       & 3       \\ \hline
\end{tabular}
\vspace{-15pt}
\label{chi_square}
\end{table}

\vspace{-10pt}
\subsubsection{Proportion of student-generated input in prompts:}
We also examined whether writing more of the prompt oneself and copy-pasting less was associated with learning gains, and whether there were differences between conditions.

In Session 2, students wrote on average $71\pm57$ characters themselves per prompt and $457\pm564$ copy-pasted characters per prompt. In Session 3, these values were $64\pm45$ typed characters and $280\pm401$ copy-pasted characters.

We implemented multiple regression models with learning as dependent variables, and the interaction between proportion of student-generated text in prompts as the independent variables on the session 3 data. We found a marginally significant positive effect of proportion of student-generated input on learning gain, $\beta=0.43, p=.075$, $95\% \ CI [–0.05, 0.91]$\footnote{The model accounts for a small portion of the variance ($\text{adj}\ R^2=.041 , p=.075$).}.

We found that students using the structured interface in session 2 had a significantly higher average of self-written text in prompts, $t(56)=2.07$, $p=.048$. However, no difference was found in session 3, $t(53)=0.48$, $p=.69$, when the normal ChatGPT interface was being used.

\vspace{-5pt}
\subsection{Students' perception of ChatGPT between conditions}
\vspace{-5pt}

The previous analysis shows that despite the lack of significant differences in learning gains or practice lab performance between conditions, students in the intervention group were more likely to engage in productive prompting behaviors while using the structured interface, but not when the structured interface was removed. To better understand why no transfer in behavior occurred, we examined the post-intervention perception survey data.

When asked about their overall perception of ChatGPT use, independent of the learning tasks, students in the intervention group were more likely to consider the use of ChatGPT to be effective for solving tasks, $U=345$, $p=.01$, Cohen's $d=0.63$, to generate high quality code, $U=387$, $p=.046$, Cohen's $d=0.52$, to be easy to use, $U=324$, $p=.0042$, Cohen's $d=0.76$, to be enjoyable, $U=334$, $0.006$, Cohen's $d=0.62$, and to continue to use ChatGPT in future practice lab sessions, $U=392$, $p=.049$, Cohen's $d=0.46$. 


However, when asked about the future use of the structured interface, 60\% of the students in the intervention group wanted to continue to use ChatGPT exclusively, compared to only 12\% who wanted to use the structured interface only, 20\% who were willing to use both, and 8\% who did not want to use any type of LLM. Students who expressed interest in using the structured interface indicated that it was ``more effective and more efficient in answering [their] questions'' (student a), with some having even (wrongly) thought that the model was ``pretrained on elements of the course'' (student b), with results that were ``more focused on the course content [thus providing] information and explanations that are more accurate'' (student c), which in turn was thought to have helped ``speed up the task completion'' (student d). One student did nevertheless mention that they learned to improve the quality of their prompts through the structured interface and would apply what they learned when using ChatGPT.

However, of the 40 justifications given for why students preferred to use ChatGPT only, we found the following reasons :
(A) Preference for ChatGPT's interface (11/40). (B) Disliking being forced to fill out multiple sections in the structured interface (2/40). (C) Did not see the added value of the structured interface compared to using ChatGPT directly (14/40). (D) Were used to using ChatGPT (9/40) or had a subscription (1/40), had access to more advanced models (3/40) or used GitHub Copilot (5/40).

In total, about 75\% of these students preferred ChatGPT only because of already having established working patterns with it, preferring its user interface and not seeing the added value of using the more constrained interface. These findings might partially explain why students who interacted with the structured interface during the interventions had a more positive perception of ChatGPT after the intervention, and also why some of the positive behaviors found in the use of the structured interface did not transfer.

\vspace{-5pt}
\subsubsection{Limitations}
\vspace{-5pt}

Given the relatively small sample size of students who ended up participating across sessions 2 and 3 (n=58), the statistical power of our analyses is limited. To strengthen our claims, we supplemented the quantitative results with qualitative analyses. Nonetheless, further research with larger samples and across diverse contexts is needed to replicate and extend these results.

\vspace{-5pt}
\section{Discussion and Conclusion}
\vspace{-5pt}

LLMs are increasingly being used by teachers and students alike, but research has indicated that not all LLM use is equivalent, with some types of behavior even being detrimental to learning. It is therefore relevant to understand how to support students to use LLMs in a way that supports learning and to help them develop good prompting behaviors for using such systems in general. We addressed this question with a mixed-methods experimental study to investigate (i) whether using a structured interface to guide prompting behaviors during a multi-session intervention contributes to improved prompting behaviors, performance, and learning, and (ii) whether potential improvements in prompting behaviors and their impact on learning transferred to subsequent ChatGPT use. Students engaged in three practice labs. In the first two practice labs, 58 students either used ChatGPT with the normal interface (control group with 32 students), or ChatGPT with the structured interface (intervention group with 26 students). In the third session, all students used the normal ChatGPT interface. The structured interface was designed based on research on effective questioning strategies in ill-structured problem-solving contexts and required that students think about the type of prompt, and structure the prompts' content before submitting to the LLM. We used pre- and post-measures of learning, process and performance data, and perception questionnaires.

\vspace{-5pt}
\subsection{Structured interfaces can promote productive prompting behaviors, but these do not transfer to unstructured interfaces}
\vspace{-5pt}
 
A first analysis of performance and learning revealed no significant differences between the conditions. To understand why, we investigated whether specific types of prompting patterns were overall more conducive to learning. Specifically, when considering the interaction between prompt type and prompt attributes, we found a positive interaction effect on learning gains between the proportion of ``Development'' prompts (prompts focused on developing code) with ``Clarity'' (proportion of prompts with clear and explicit requests), and between overall proportions of ``Clarity'' and ``Understanding'' (prompts focused on understanding-related questions). These findings align with prior work on student-generated questions that shows that students' learning improves when they articulate their ideas and questions more clearly \cite{byun2014relative}.  

We found no differences between conditions in either prompt attributes or types when examining them separately. But we found that students in the intervention group had a higher proportion of prompts of type ``Development'' with the attribute ``Clarity'', i.e. a higher proportion of prompts that were found to be productive for learning, while using the structured interface. 

This aligns with Kumar et al. \cite{kumar2023impact} who observed that pedagogically informed guidance, such as metacognitive questioning, reduced superficial interactions and encouraged deeper engagement with learning content. These exploratory results support the view that the impact of LLMs on learning outcomes depends on \textit{how} they are used \cite{lehmann2025aimeetsclassroomlarge}. Unfortunately, and as found in another recent study \cite{bastani2024generative}, the changes in prompting behavior did not extend beyond the intervention: once the students were free to use ChatGPT, they tended to have the same prompting behaviors to those in the control group. 


Similarly, we observed a trend where the proportion of student-generated text in prompts (in contrast to the proportion of copy-pasted prompts) correlated with better outcomes. While additional data is required to substantiate this effect, the pattern aligns with the hypothesis that personalized, self-generated text in interaction with AI tools can support learning \cite{kumar2023impact}. Students using the structured interface tended to have higher proportion of self-generated input in their prompts, but as before, this effect disappeared once guidance was removed. 

\vspace{-5pt}
\subsection{Why don't ``good'' prompting practices persist when scaffolding is removed?}
\vspace{-5pt}

A final analysis of the student survey data revealed that participants using the structured interface perceived the use of ChatGPT overall (independent of the structured interface) as significantly more positive than students in the control group. While this might seem surprising, it appears to be a reaction to the structured interface, which many students considered to be overly constraining: less than 20\% of students in the intervention group appreciated the added value of the structured interface. In contrast, about 75\% of the students in the intervention group did not see an added value in this interface compared to the normal user interface, with habit of use of ChatGPT and the more appealing user interface indicated as reasons. 

This indicates that when trying to teach good prompting behaviors through the use of novel pedagogical LLM tools, or ``pedagogy-fied'' interfaces for existing LLMs, we need to consider students' prior experiences and habits with LLMs. Even if ChatGPT has been in broad use for only about 2 years by now, a majority of students have already had significant experiences with this tool, and thus developed their own habits and expectations that might increase the resistance to adopting novel other tools. Thus, if we are to design effective pedagogical approaches to help students develop good prompting behaviors, we need to take their history with LLM tools into consideration, as well as their expectations of such tools. Two intervention sessions were not sufficient for the students to adopt new behaviors. Effective interventions likely need a longer duration, and the integration of more explicit approaches, such as elaborating productive prompting behaviors or addressing possible contrasts between pre-existing tools and these more ``pedagogical'' tools \cite{choi2025survey}. 

\vspace{-5pt}
\subsection{Conclusion and future work}
\vspace{-5pt}

Overall the findings confirm ``the complexities of designing learner–LLM interactions'' \cite{kumar2023impact} to support learning, and to develop productive prompting behaviors. This work underscores the challenge of transfer also in this context \cite{barnett2002and}: one-off scaffolding or training often fails to translate into long-term skill adoption. Learners frequently abandon strategies when external support vanishes \cite{bastani2024generative}, especially if those strategies require extra effort \cite{sweller1998cognitive}. 
Future research should therefore explore strategies to maintain effective AI interaction behaviors in educational contexts beyond structured interventions.
For instance, structured interfaces could integrate more personalization, such as adaptive scaffolding or reflective follow-up questions, may further engage the user and support metacognitive reflection. Moreover, combining interface support with explicit instruction and classroom dialogue around effective LLM use could help learners better understand and retain productive strategies. 
Ultimately, fostering lasting changes in students’ interactions with LLMs may require not only interface-level nudges, but also sustained pedagogical and motivational support that aligns with learners’ goals to promote durable learning outcomes.